\documentclass[sigconf,nonacm]{acmart}

\usepackage{booktabs}
\usepackage{adjustbox}
\usepackage[ampersand]{easylist}
\usepackage{array}
\usepackage{multirow}
\usepackage{makecell}

\newcommand{\pp}{\textit{$P_1$-$P_2$}~}
\newcommand{\pandp}{\textit{$P_1\!\And\!P_2$}}

\AtBeginDocument{%
  \providecommand\BibTeX{{%
    \normalfont B\kern-0.5em{\scshape i\kern-0.25em b}\kern-0.8em\TeX}}}

\setcopyright{acmcopyright}
\setcopyright{none}
\copyrightyear{2018}
\acmYear{2018}
\acmDOI{10.1145/1122445.1122456}

\acmConference[WSDM '21]{Woodstock '18: ACM Symposium on Neural
  Gaze Detection}{June 03--05, 2018}{Woodstock, NY}
\acmBooktitle{Woodstock '18: ACM Symposium on Neural Gaze Detection,
  June 03--05, 2018, Woodstock, NY}
\acmPrice{15.00}
\acmISBN{978-1-4503-XXXX-X/18/06}

\begin{document}

\title{On the impact of predicate complexity in crowdsourced classification tasks}
\titlenote{This is a post-peer-review, pre-copyedit version of an article accepted to the 14th ACM International Conference on Web Search and Data Mining, WSDM 2021.}

\author{Jorge Ram\'irez$^{+}$, Marcos Baez$^{\Diamond}$, Fabio Casati$^{\dagger}$, Luca Cernuzzi$^{\circ}$, Boualem Benatallah$^{\triangle}$,}
\author{Ekaterina A. Taran$^{\ddagger}$, and Veronika A. Malanina$^{\ddagger}$}

\affiliation{
  \institution{$^{+}$University of Trento, Italy. $^{\Diamond}$Université Claude Bernard Lyon 1, France. $^{\dagger}$ServiceNow, USA.}
}
\affiliation{
  \institution{$^{\circ}$Catholic University Nuestra Señora de la Asunci\'on, Paraguay. $^{\triangle}$University of New South Wales, Australia. $^{\ddagger}$Tomsk Polytechnic University, Russia.}
}
  
\definecolor{royalblue(web)}{rgb}{0.25, 0.41, 0.88}
\newcommand{\pending}[1]{\textcolor{royalblue(web)}{#1}}

\renewcommand{\shortauthors}{Ram\'irez, et al.}


\begin{abstract}
This paper explores and offers guidance on a specific and relevant problem in task design for crowdsourcing: how to formulate a complex question used to classify a set of items. In micro-task markets, classification is still among the most popular tasks. We situate our work in the context of information retrieval and multi-predicate classification, i.e., classifying a set of items based on a set of conditions. 
Our experiments cover a wide range of tasks and domains, and also consider crowd workers alone and in tandem with machine learning classifiers.
We provide empirical evidence into how the resulting classification performance is affected by different predicate formulation strategies, emphasizing the importance of predicate formulation as a task design dimension in crowdsourcing.
\end{abstract}





\maketitle


\section{Background \& Motivation}

Micro-task crowdsourcing today is still an art. Indeed, it is not surprising that companies charge hefty consulting fees to help businesses set up and run crowdsourcing tasks. 
Successful projects involve designing and harmonizing several aspects, from designing the user experience to task design, training and test settings, and to seemingly easy problems such as how to ask questions and elicit truthful, accurate answers \cite{DBLP:journals/csur/DanielKCBA18} --- all while meeting budget constraints and treating your workforce fairly and with respect.
%
%
%

For example, longer instructions affect the task uptake by workers by three times, while showing concrete solution examples improve accuracy up to ten times \cite{DBLP:conf/hcomp/WuQ17}, depending on the type of task. Mechanisms to combat task spammers are often essential, since without them a task can easily get half of the answer as invalid even on simple tasks, although it raises the contributors' efforts (and cost)
\cite{DBLP:conf/chi/KitturCS08}.
Budget is also a limiting factor, and reward strategies and optimization can also affect the results \cite{DBLP:conf/cscw/ChengB15,DBLP:journals/pacmhci/CallaghanGMLL18,DBLP:journals/jamia/WallaceNMCST17,KrivosheevCSCW2018}.
The list is almost endless, so much that it is motivating crowdsourcing researchers to prepare design and reporting guidelines for crowd experiments \cite{Ramirez2020DrecCSCW}. 

This paper explores and provides guidance on a specific but important aspect of crowdsourcing task design: how to ask "complex" questions to the crowd to classify items. 
Classification in general is by far the most popular type of crowdsourcing tasks\footnote{
A relatively recent worker survey on Appen, previously Figure Eight, shows that 45\% of jobs are classification tasks \cite{DBLP:conf/ht/GadirajuKD14}. Also, 60\% of the builtin templates offered by Amazon Mechanical Turk constitute classification tasks, and 40\% in Yandex Toloka.
}. 
In this paper we study classification in the context of information retrieval and multi-predicate classification problems, that is, tasks where the crowd has to select items that meet a set of conditions. The "complexity" of the question comes therefore from the fact that it is \textit{composite}, and we want our crowd worker to state if items satisfy our set of conditions (predicates). 
This is a very common task we do implicitly or explicitly countless of times in our daily life and that often appears in crowd tasks as well (from selecting hotels that have certain characteristics of interest \cite{DBLP:conf/hcomp/LanRST17} to screening papers for systematic literature reviews \cite{DBLP:journals/jamia/WallaceNMCST17}). 
Indeed, any conjunctive query is an instance of such problem and abundant prior research on crowd query processing studied how to efficiently retrieve items from a potentially large set \cite{DBLP:conf/sigmod/FranklinKKRX11,DBLP:conf/sigmod/ParameswaranGPPRW12,DBLP:journals/pvldb/ParkW13}.

We tackle this problem because it is common enough to be of widespread interest and nuanced enough (as we show in this paper) to require a detailed investigation, and it can be framed so that it can result in reusable knowledge for task designers. In particular, we set to study the following research question:


\textit{How does the way we ask a composite question impacts the individual and aggregate performance of crowd workers?}



We investigate the question both in the context of crowd-only classification and in \textit{hybrid classification}, an increasingly common approach where humans and machines work together to solve a classification problem. We analyze the problem based on both characteristics of the question and of the task, such as task "length" (e.g., length of the document to read for text classification tasks), task domain, task difficulty, and class balance. 
Surprisingly, the crowdsourcing literature somewhat overlooked predicate complexity in classification tasks. 
First, complex predicates may require longer task instructions, which is known to correlate positively with the perceived complexity (as seen by workers \cite{YangHCOMP2016}), impact task intake (most workers tend to quit after inspecting the instructions \cite{DBLP:conf/wsdm/HanRGSCMD19}), and, therefore, the latency.
Second, increased task complexity naturally demands more effort from workers, challenging accurate and fair compensation \cite{Whiting2019FairWC}.
Last, task complexity plays an important role in the quality of the results obtained \cite{DBLP:conf/chi/ChengTIB15,DBLP:conf/hcomp/KrauseK15}.

The main contributions of this work are as follows. 
We introduce the problem of predicate formulation for crowd classification tasks as a relevant design dimension (to enhancing worker performance and enabling Human-AI collaboration). 
We study complexity in classification problems on a broad landscape of tasks considering categorization and classification, verification, content moderation, and sentiment analysis tasks (see \cite{DBLP:conf/ht/GadirajuKD14} for a taxonomy of task types in crowdsourcing). Our experiments, therefore, cover multiple domains and leverages human and machine classifiers.
We provide empirical evidence on the impact of predicate formulation on classification outcomes, suggesting performance gains when querying complex predicates as multiple simpler questions.
We also provide insights into the expected performance of different formulation strategies under different i) problem settings such as predicate selectivity and class distribution, and ii) task design choices such as querying predicates on the same or separate tasks.
The experiments also offer preliminary evidence on the potential of predicate formulation in the context of hybrid classification, suggesting performance gains even in its simplest collaborative approach, by assigning crowd and machines parts of a complex predicate they are more suited to classify.
%
%
Last but not least, we contribute datasets derived from our experiments\footnote{\url{https://github.com/TrentoCrowdAI/simpler-predicates}}.
\section{Related Work} 

\textbf{Task design in crowdsourcing}

\noindent Task design is a multi-dimensional problem with a rich body of work in the crowdsourcing literature \cite{DBLP:journals/pvldb/JainSPW17}. 
"Design" does not only mean the actual task interface, but also the mechanisms to deploy, coordinate, and assign tasks to workers, the tools to assure high-quality contributions, and budget management~\cite{DBLP:journals/csur/DanielKCBA18}.
The lessons learned from this literature spawn on best practices for designing effective tasks (given the impact task design has on the resulting performance), and methods for performing crowdsourcing studies.


Crowdsourcing results are sensitive to subtle changes in task design.
Poor instructions may lead workers to misinterpret the task and produce subpar responses \cite{DBLP:conf/hcomp/WuQ17}. The clarity of the task \cite{DBLP:conf/ht/GadirajuYB17} and how it is framed (whether meaningfully or not) \cite{CHANDLER2013Breaking} may also swing workers' performance.
The prevalence of malicious workers in platforms asks for design decisions that account for this and guard quality (e.g., equip tasks with mechanisms to combat spammers \cite{DBLP:conf/chi/KitturCS08}).
Similarly, task design could aid worker performance, in the form of assistance to workers \cite{Wilson2016WWW,ramirez2019}, proper compensation for effort-intensive tasks \cite{DBLP:conf/www/HoSSV15}, or by rigorous training protocols \cite{DBLP:conf/naacl/LiuSBLLW16} and feedback loops \cite{DBLP:conf/cscw/DowKKH12}.
Latency also matters and can be affected by ineffective instructions causing task abandonment \cite{DBLP:conf/wsdm/HanRGSCMD19} or generating mistrust in task requesters \cite{DBLP:conf/cscw/KitturNBGSZLH13}. 
However, fair compensation can help to speed up task intake and how much workers contribute \cite{DBLP:conf/www/HoSSV15}.

These lessons provided valuable insight into properly designing and running crowdsourcing studies. As design choices may swing the results obtained, it can also affect the validity of experimental outcomes \cite{DBLP:conf/chi/KitturCS08}.
Choices in task design can amplify biases inherent to crowdsourcing environments. Task clarity influences how workers pick tasks and, therefore, introduce selection effects \cite{DBLP:conf/ht/GadirajuYB17}. The active pool of workers varies as hours go by \cite{DBLP:conf/wsdm/DifallahFI18}, and with this, different decisions affecting when a crowdsourcing job runs could result in unanticipated performance differences and confounding factors \cite{Qarout2019PlatformRelatedFI}.
The lack of built-in support from crowdsourcing platforms makes it difficult to run controlled experiments, making simple between-subjects design a challenging endeavor \cite{DBLP:conf/chi/KitturCS08}. A common approach involves identifying workers via browser fingerprinting  \cite{gadiraju2017improving} and then using an external server to randomize participants to experimental conditions \cite{ramirez2019}. 
These challenges motivated the research community towards developing guidelines for designing and reporting crowdsourcing experiments \cite{Porter2020,Ramirez2020DrecCSCW}.



\smallskip\noindent\textbf{Multi-predicate classification}


\noindent We study predicate formulation in the context of problems regarded as \textit{finite pool} classification \cite{DBLP:conf/hcomp/NguyenWL15}, where we have a finite set of items to classify according to a set of criteria (potentially) unique to the problem.
Systematic literature reviews are one instance of this problem, and have been heavily-studied in the crowdsourcing literature \cite{Mortensen2016crowd,KrivosheevHcomp2017, DBLP:journals/corr/SunCWLLMW16,Weiss2016CrowdsourcingLR}. Mortensen and colleagues \cite{Mortensen2016crowd} tested the feasibility of leveraging crowdsourcing, given the costs associated with producing SLRs \cite{DBLP:journals/jamia/WallaceNMCST17}. They found that task design plays a major role in the quality of the results, as well as this can vary from predicate to predicate.
Krivoshev et al. \cite{KrivosheevHcomp2017} proposed models and algorithms to crowdsource SLRs, offering quality and budget trade-offs to guide how to invest in the crowdsourcing tasks.
Budget limits entire crowdsourced solutions, works have also focused on leveraging machine classifiers in tandem with crowd workers \cite{DBLP:journals/jamia/WallaceNMCST17,KrivosheevCSCW2018}. For example, leveraging strategies such as classifying ``easy'' items first with ML and crowd for the rest \cite{DBLP:journals/jamia/WallaceNMCST17} or modeling tasks and workers to determine promising predicates to filter out items.

Multi-predicate classification is also studied in the context of information retrieval. 
A common problem is to determine an optimal order of the predicates (to query the crowd for labels) to filter out tuples \cite{DBLP:conf/sigmod/ParameswaranGPPRW12,DBLP:conf/hcomp/LanRST17,DBLP:conf/cikm/RekatsinasDP19,DBLP:conf/cikm/WengLHF17}. 
Similarly, work in crowd-powered databases studied how to leverage crowdsourcing to extend the capabilities of database systems to answer complex multipart queries over flexible (or on-demand) schemas \cite{DBLP:conf/sigmod/FranklinKKRX11,DBLP:journals/pvldb/ParkW13}.

Despite the vast body of work on task design and on information retrieval / multi-predicate classification, to the best of our knowledge, we are the first to study the impact on how the (complex) information retrieval question is formulated, a dimension that affects all of the prior art.
Our experiments, over an ample range of tasks, emphasize the importance of the predicate formulation as a problem, and show its impact on classification outcomes.

\section{Problem and Approach}
We now define and scope the crowdsourced classification problem, and in Section \ref{sec:hybrid}, we introduce the crowd-machine variant.


The task we seek crowd help for is to identify all items in a set $I$ that meet a complex predicate $\mathcal{P}$, defined as the conjunction of predicates $\{p_1,p_2,\dots,p_n\}$.
For example, taking a common problem from the literature (screening scientific papers \cite{KrivosheevHcomp2017,DBLP:journals/jamia/WallaceNMCST17})
, $I$ could be a set of scientific articles returned by a keyword-based query on Scopus, and we may seek papers reporting experiments on older adults living in Africa ($\mathcal{P} = p_1 \land p_2$, where $p_1$: \textit{``Is the study population 65+ years?"} and $p_2$: \textit{``Is the population living in Africa?"}). 
%
To solve this problem, we have to our disposal a set of crowd workers $W$, a budget $B$, and a quality goal (or loss function) $L$ to meet.

The predicate formulation problem seeks to determine how to ask the question in the context of multi-predicate classification. There are different ways to formulate a complex predicate, and, in this paper, we study specifically three ways: i) ask the complex question (e.g., \textit{``Is the study on 65+ years old adults living in Africa?"}), ii) break the composite question into component predicates, but ask them as part of the same task, and iii) make each predicate a task of its own (which also means that a crowd worker only sees one predicates and assesses many items for it).
%



%
We approach this problem systematically, considering both characteristics of the question and tasks. 
To give breadth to our analysis, we explore a broad landscape of tasks (categorization and classification, verification, content moderation, and sentiment analysis tasks) representing different domains and task difficulty levels.
We focus our experiments on document retrieval (text classification) and consider documents of different lengths, given the associated effort incurred on workers to (understand and) assess text, and the potential influence of the documents' length on performance \cite{DBLP:conf/chi/ChengTIB15,ramirez2019}.
Our focus on text stems from the fact that it is a recurrent use case in the literature \cite{DBLP:conf/www/HoSSV15,Wilson2016WWW,KrivosheevHcomp2017}, and annotating images are deemed simpler in comparison to annotating texts \cite{DBLP:conf/hcomp/KrauseK15}.
Finally, crowdsourcing tasks are prone to worker biases \cite{DBLP:conf/hcomp/FaltingsJPT14}, which could be caused by frequently assigning items to the same class. Therefore, we consider different class distribution scenarios to study the predicate formulation problem in crowdsourcing contexts.

\section{Crowdsourcing Experiment}

This experiment studies the impact of the task design alternatives on the performance of crowd workers.
We focus on the simpler case where a complex predicate is composed of two simpler ones.
%
We show the individual and collective performance gains related to predicate reformulation, and how the nature of the problem influences the resulting performance.

\smallskip

\subsubsection*{Datasets}~\label{sec:dataset}
We considered datasets with different characteristics in terms of domain, predicates, document length, and difficulty (classification accuracy), in line with prior art \cite{ramirez2019,KrivosheevCSCW2018}. 
%
%
The datasets come from systematic literature reviews (SLRs), customer feedback analysis, content moderation and crowd verification, and are representative of multi-predicate screening problems from the literature \cite{KrivosheevCSCW2018,RamirezBMC2019,WikiDetoxDataset}. 
%
See the supplementary material\footnote{\url{https://tinyurl.com/simpler-predicates-supp}} for details on the predicate composition for each of the reference datasets.

\textbf{Virtual reality exergames}. This dataset was produced and annotated by the authors as part of their  investigation into overlaps between SLRs. 
We identified a pool of $80K+$ scientific articles from multiple SLRs that share some predicates.
%
%
From this pool of papers, we built the \textit{Exergame-VR} dataset that consists of $500$ articles from $4$ SLRs with high overlap in terms of predicates and papers within their scope. 
Additionally, we split the documents into two buckets based on their length: \textit{short} ($150$ items with length  $\leq 230$ words) and \textit{long} ($350$ items with length $> 230$ words).

\textbf{Amazon product reviews}. This dataset contains 100k reviews of products that are sold in Amazon \cite{KrivosheevCSCW2018}. It is labeled according to the following two predicates: \textit{1. Book:} \textit{``Is it a book review?"}, and \textit{2. Negative:} \textit{``Is it a negative review?"}. 
We randomly selected $236$ reviews ($118$ \textit{short}, and $118$ \textit{long}) to create the \textit{AMZ-reviews} dataset.

\textbf{Wikipedia detox}. This dataset from Wulczyn et al. \cite{WikiDetoxDataset} contains 100k comments from ``Talk pages" in Wikipedia, labeled by crowd workers on whether each of the comments contains a personal attack (or an attack of another kind). From this pool of 100k items, we built \textit{Content-Moderation}, a dataset of $118$ \textit{long} documents (comments with $> 230$ words) labeled on two predicates. 

\textbf{Verifying crowd contributions}. In \cite{RamirezBMC2019}, the authors contribute datasets where workers provided a binary label to a relevance question, and a highlighted excerpt to justify the labeling.
We built the \textit{Verification} dataset based on \cite{RamirezBMC2019}, selecting 118 \textit{long} documents labeled according to predicates that determine whether the judgment and highlighted passage are correct.
%
These tasks are relevant to iterative workflows, where workers act as reviewers \cite{DBLP:conf/uist/LittleCGM10}.

\textbf{Economic inequalities in older adults}. This dataset is part of an ongoing SLR on assessing the inequalities in older adults. It contains 2619 papers. From this pool of documents, we selected and labeled $151$ items to build \textit{Inequality-OA}, a dataset of \textit{long} abstracts.

\subsubsection*{Design}
\begin{figure}[h]
  \centering    
  \includegraphics[width=\columnwidth]{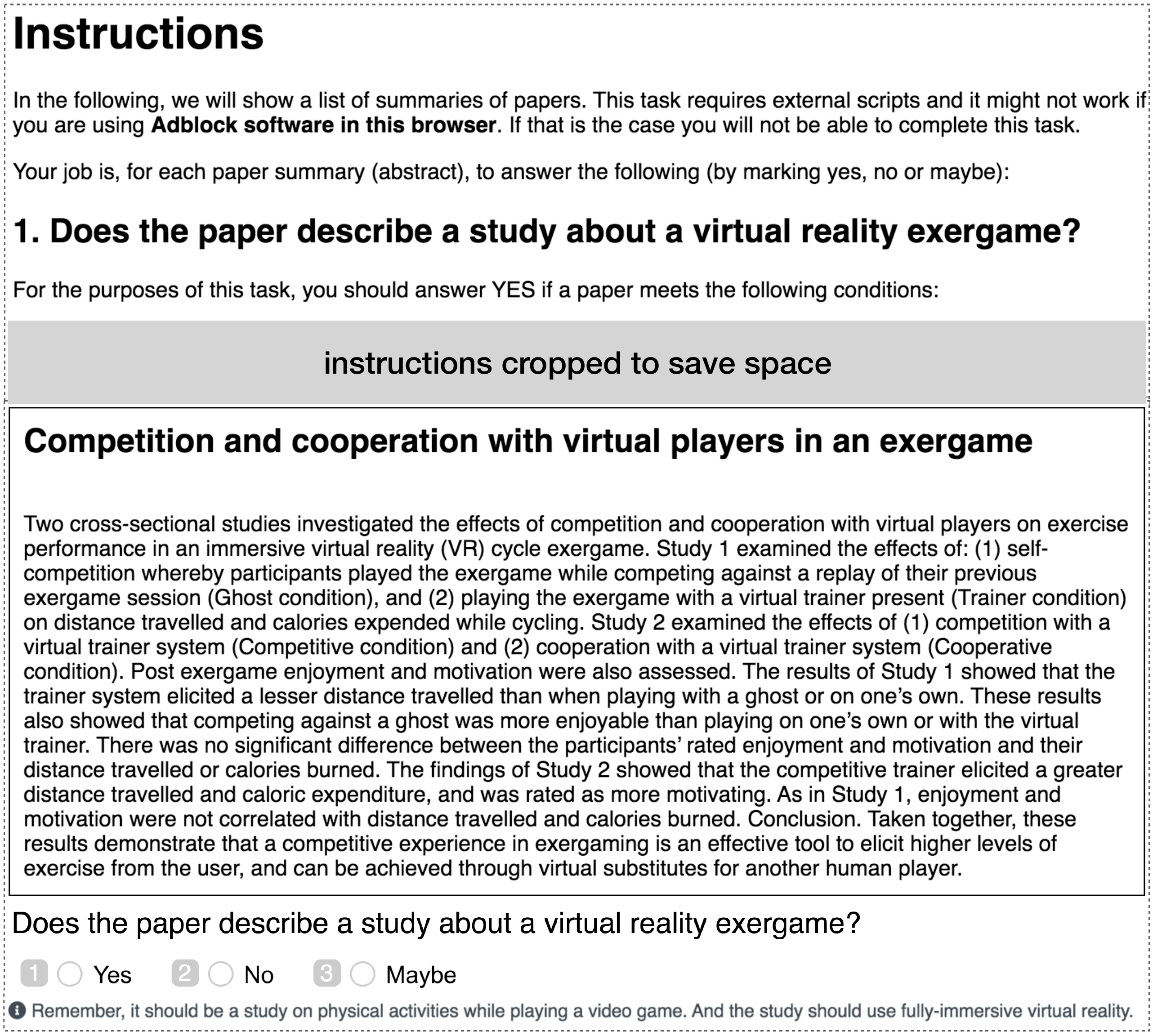}
  \caption{\textmd{The task interface used in the crowdsourcing experiments. The interface shows the complex predicate $\mathcal{P} = p_1 \land p_2$ for the \textit{Exergame-VR} dataset, with \textit{$p_1$:} \textit{``Does the paper describe a study that uses an exergame?"}, and \textit{$p_2$:} \textit{``Does the paper describe a study that uses virtual reality for physical training?"}. }}~\label{fig:task-interface}
\end{figure}

The task performed by workers in our experiment consisted of reading a piece of text and answering one or two binary questions of different complexity levels depending on the task design. Figure \ref{fig:task-interface} shows an example of a task (inspired by prior art \cite{KrivosheevHcomp2017,CrowdRev2018}).

We selected $118$ items per dataset, reserving $18$ for training workers (training items), and $100$ for the actual task, where 34 of these items were used for quality control (control items).
%
We consider two scenarios for the class distribution in these datasets: 60-40 and 80-20.
In the \textit{60-40} case, we selected items in each dataset according to a distribution of roughly 40\% included (\textit{IN}), 60\% excluded (\textit{OUT}). 
Included means that the documents satisfy all predicates $p_j \in \mathcal{P}$ for a given dataset (i.e., documents have a value of $1$ for the predicates that constitute $\mathcal{P}$). Excluded documents are those that satisfy only one of the predicates or none of them.
The excluded documents we distributed equally, whenever possible, between the three exclusion cases\footnote{Representing the two predicates in each dataset as $p_1$ and $p_2$, the three exclusion cases are 1) $p_1=1$, $p_2=0$; 2) $p_1=0$, $p_2=1$; 3) $p_1=0$, $p_2=0$.}. 
As the name suggests, the \textit{80-20} case represents a setting with roughly 20\% of items included and 80\% excluded.
This skewed setting tends to be problematic in crowdsourcing since it may bias workers towards the most frequent answer \cite{DBLP:conf/hcomp/FaltingsJPT14}. 
For this reason, and quality control purposes, the training and control items follows a 30~\textit{IN} and 70~\textit{OUT} distribution, making sure that each page of work shows items from both classes.
%


We consider four experimental conditions for our crowdsourcing experiments, each condition represents a variation of the task interface shown in Figure \ref{fig:task-interface}. 
The \textit{baseline} condition we use as control, and it asks workers a complex predicate $\mathcal{P}$ (a question that integrates both predicates in a dataset, as indicated in \ref{fig:task-interface}). 
%
The \pp~condition represents the task alternative that asks the constituents of $\mathcal{P}$ on the same task.
The conditions \textit{$P_1$} and \textit{$P_2$} represent tasks that asks workers only one simpler predicate (predicates $p_1$ and $p_2$, respectively).

Initially, we considered both \textit{short} and \textit{long} documents. 
However, in a pilot study, we observed that the task alternatives did not improve over the \textit{baseline} when considering \textit{short} documents, suggesting that these may be more suitable for tasks where workers face longer documents \cite{Wilson2016WWW,ramirez2019}. Therefore, we consider only \textit{long} documents in our study.

We followed a between-subject design, assigning workers to one of the four experimental conditions. Workers judged a maximum of $18$ documents that we divided into $6$ pages of $3$ items per page ($6\times3$ design), and we only allowed workers that understand English.
We required workers to perform a training task (one page of work) before advancing to the main task, a quality control mechanism typically done in crowdsourcing research \cite{DBLP:conf/chi/MitraHG15}. 
Workers that scored $100\%$ advanced to the main task, where we included control items as an additional quality assurance mechanism. We required workers to maintain an accuracy level of 100\% for the \textit{AMZ-reviews} dataset and  76\% for the rest of the datasets\footnote{Prior art \cite{KrivosheevCSCW2018} shows that the baseline performance was quite high for \textit{AMZ-reviews}; therefore, we defined the 100\% quality threshold for this dataset.}. We paid workers between \$$0.09$ and \$$0.21$ per page of work (depending on the condition and dataset), aiming at an hourly rate of $7.5$ USD.
We collected contributions from workers on the Yandex Toloka platform\footnote{\url{https://toloka.yandex.com/}}, asking $3$ votes per item in the datasets.
We defined a timeframe from 14:00 to 21:00 GMT+1 for running the experiment, running each dataset separately with a time gap between these.  We executed each of the experimental conditions in parallel and balanced the contributions from each geographical bucket (\textasciitilde$33\%$ per bucket within each condition).
%


We inspected the demographics of Toloka and noticed that roughly 90\% of workers come from Russian-speaking countries, where Russia and Ukraine contribute the majority of the workers (${\sim}79\%$ and ${\sim}10\%$ respectively). Besides only allowing workers that understand English, we decided to create $3$ geographic buckets: Russia, Ukraine, and the ``Rest of the world", balancing the contributions from these buckets in our experiments to avoid any bias due to demographics.

We used an external server to assign workers to experimental conditions in a round-robin fashion, blocking workers from jumping between conditions (to avoid learning effect). We added a custom JavaScript code to the task interface to call the external server and render the experimental condition accordingly \cite{CrowdHub2019}.



\subsection{Results}

We collected a total of $8250$ judgments from $1185$ workers across the datasets we considered in this experiment. 
Here we describe our results to determine the impact of asking the complex predicate $\mathcal{P}$ vs. leveraging its simpler constituents on the classification performance of crowd workers.



\smallskip\noindent\textbf{Worker accuracy}\smallskip

\noindent We use the ground-truth labels available in the datasets to determine the classification accuracy of workers in each of the experimental conditions.
The median accuracy of workers in the \textit{baseline} was $0.89$ for AMZ-reviews and $0.67$ for the rest (Exergame-VR, Content-moderation, Verification, and Inequality-OA), it would seem that workers found it easier to judge product reviews than documents from the rest of the domains.

\begin{figure}[h]
  \centering    
  \includegraphics[width=0.9\columnwidth]{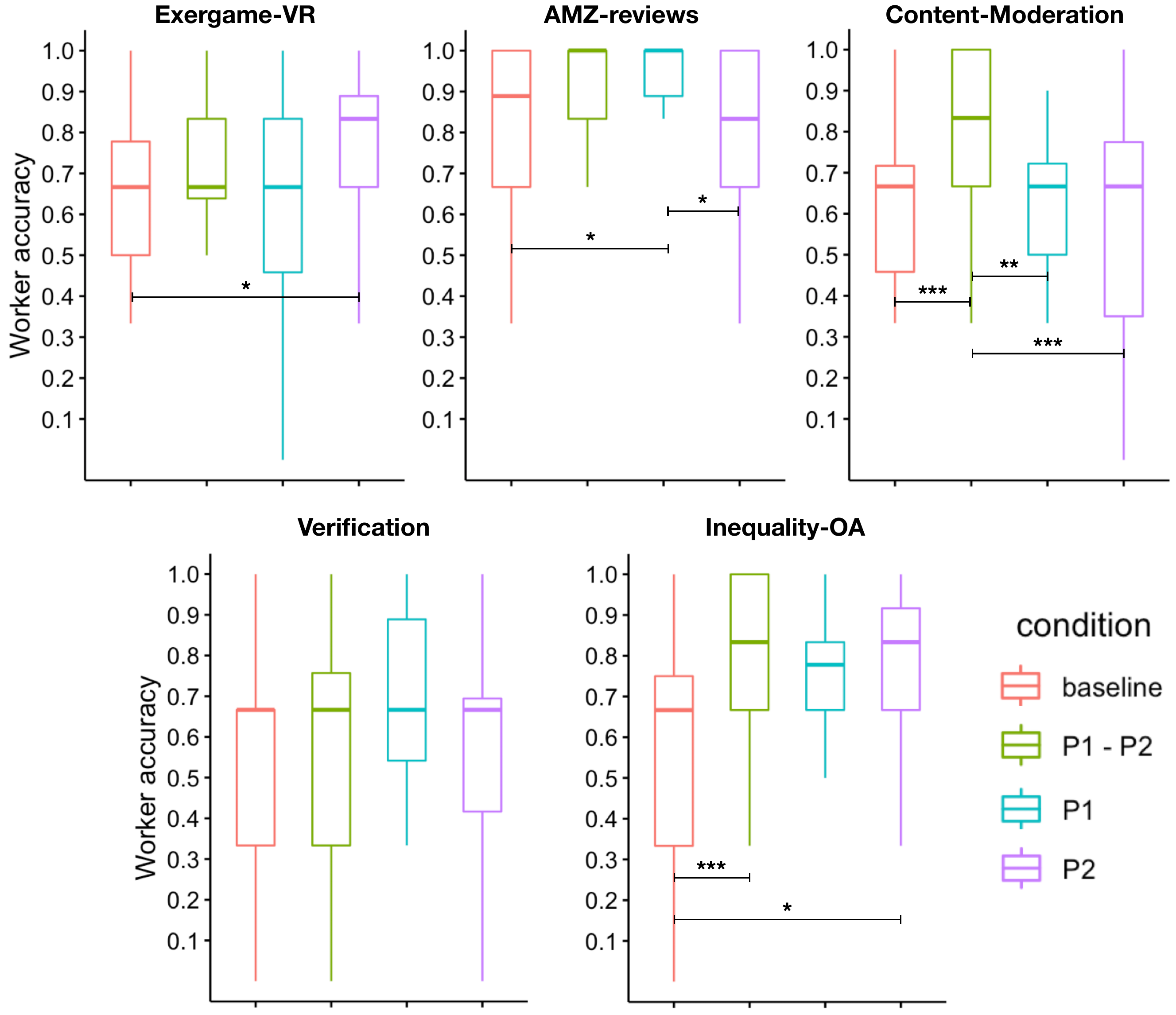}
  \caption{\textmd{Worker accuracy by experimental condition for the \textit{60-40} case. The lines indicate significant differences, coding p-values as *: $p \leq 0.05$, **: $p \leq 0.01$, ***: $p \leq 0.001$, ****: $p \leq 0.0001$.}}~\label{fig:acc-balanced}
\end{figure}

We test the significance of the difference in worker accuracy using the \textit{Kruskal-Wallis H test}\footnote{In our pilot study, we noticed that the observations do not follow a normal distribution}. 
The test indicates that there is a significant difference between the experimental conditions in $4$ out of $5$ datasets ($p < 0.05$ for Exergame-VR, and $p < 0.01$ for the rest). The results are depicted in Figure \ref{fig:acc-balanced}. 
We analyze all possible pairwise comparisons using the Dunn's Test of Multiple Comparisons \cite{DunnTest1964}, using Benjamini-Hochberg correction to reduce the probability of Type I error. 
It can be noted that either $P_1$ or $P_2$ has a significant improvement over the \textit{baseline} (3 out of 5 datasets). And that the \pp~condition significantly outperforms the \textit{baseline} in the Content-moderation and Inequality-OA datasets. 



In the skewed class distribution scenario, the \textit{80-20} case, the median worker accuracy in the baseline condition was $0.67$ for both Exergame-VR and Inequality-OA (figure omitted due to space limitations). The Kruskal-Wallis test shows no significant results between the experimental conditions (though, there is an interesting advantage of the \textit{$P_1$-$P_2$} condition where the median worker accuracy was $0.83$ in both datasets).

%
%
For our predicate formulation problem, these results suggest that by asking simpler predicates instead of a complex question, we are likely to see an increase in worker accuracy in at least one of the simpler predicate.
Furthermore, by asking more granular and simpler predicates we obtain valuable detailed information about crowd and task characteristics. 
For example, according to our results, workers were better at evaluating if a review was about a book than whether it was a negative review (simple verification vs. sentiment analysis tasks), suggesting different difficulty levels. This detailed information could equip crowd-machine algorithms to make better decisions about what to crowdsource and what to automate.
%



\smallskip\noindent\textbf{Classification performance}









\noindent We analyze the results from a collective perspective and evaluate the impact of predicate formulation in the resulting classification.

The overall classification, for a given $\mathcal{P} = \{p_1, p_2\}$ in our datasets, is derived from the conjunction of the aggregated results from each of the simpler predicates (i.e., $p_1 \land p_2$ for each item $i \in I$).
We use majority voting to aggregate the contributions from multiple workers and the $F_1$ score to assess the classification quality.
The \textit{baseline} condition already provides classification on $\mathcal{P}$ since the simpler predicates are combined in a single question, and for the \pp condition we simply take the conjunction of $\{p_1,p_2\}$. 
To make the results from $P_1$ and $P_2$ comparable to previous conditions, we introduce \pandp, which also takes the conjunction of each simpler predicate.
%
%
To compute the $F_1$ score for 
these conditions we use as ground-truth label the conjunction of the simpler predicates.

%
Table \ref{table:f1-crowd} summarizes the classification performance for the experimental conditions across our five datasets for both \textit{60-40} and \textit{80-20} cases.  The \textit{baseline} performance ranges between $0.6$ (Exergame-VR) and $0.909$ (AMZ-reviews).

We compared the performance of asking $\mathcal{P}$ directly vs. asking the simpler predicates first and then combining the results (\pp~and \pandp).
It can be observed that the conditions \pp~and \pandp, outperformed the baseline condition but not consistently across all datasets.
\pp~ outperformed the baseline in $3$ out $5$ datasets (Content-moderation, Verification, and Inequality-OA), with an increase in performance of up to 18\%. \pandp~improved over the baseline in $2$ out $5$ datasets, with an increase of up to 9\%.
In the \textit{80-20} case, the \pp~condition showed superior performance when compared to the \textit{baseline}, with an increase of up to $27\%$ (while the \pandp~fell behind the baseline).

%
We also compared the simpler predicates against the complex one. 
%
%
We observed superior classification results when formulating a composite predicate as multiple (more straightforward) questions leveraged on the same or separate tasks, even when votes are aggregated with simple majority voting.
%
Asking two simple questions on the same task (the \pp~condition) resulted in performance gains ranging from $6\%$ to $48\%$.
And the conditions $P_1$ and $P_2$ that asked a simple question surpassed the \textit{baseline} performance in all datasets, with an increase in $F_1$ score ranging from $2\%$ and up to $47\%$
In the \textit{80-20} case the $F_1$ scores in the \textit{baseline} were $0.571$ for Exergame-VR, and $0.476$ for Inequality-OA. 
In both datasets, task formulating simple predicates outperformed complex ones (when delivered separately or together on the same task), with an increase in classification performance of up to 97\%.

A closer look into the performance on the complex predicates  (\textit{baseline} condition) across the two class distribution scenarios showed that overly skewed datasets may hurt the classification performance of the crowd --- $F_1$ decreased 4\% for Exergame-VR, and 31\% for Inequality-OA.
%
%
While by leveraging simple predicates, the classification performance could remain roughly the same, except for the unusual case of $P1$ for Inequality-OA, where the performance decreased 20\%.
We believe the selectivity of $P_1$ (see our supplementary material) played a role in this drop in performance since it is equal to $0.44$ in the \textit{60-40} version of Inequality-OA and $0.20$ in the more skewed variant (also observed in the baseline).

In summary, there is evidence suggesting that complex multipart questions may benefit from disentanglement into simpler elements. As we observed, performance boosts can be obtained by formulating and presenting complex predicates as simple and more granular questions and combining back the results. 
%
However, there is no clear pattern for when each task design alternative (presenting simple predicates on the same- or separate tasks) will be the appropriate one to implement, an interesting direction for future work.


\begin{table*}
\caption{\textmd{\label{table:f1-crowd}Classification performance ($F_1$ scores) by experimental condition from the crowdsourcing experiment.}}
\centering
\begin{adjustbox}{width=.8\textwidth}
\fontsize{0.5em}{0.6em}\selectfont
\begin{tabular}[t]{lrrrrr}
\toprule
\textbf{Condition} & \textbf{Exergame-VR} & \textbf{Inequality-OA} & \textbf{AMZ-reviews} & \textbf{Wiki-detox} & \textbf{Verification}\\

 \textit{Distribution} & \textit{60-40 (80-20)} & \textit{60-40 (80-20)} & \textit{60-40}& \textit{60-40} & \textit{60-40}\\
 \midrule
\hspace{1em}Baseline & 0.600 (0.571) & 0.691 (0.476) & 0.909 & 0.697 & 0.674\\
\hspace{1em}P1 - P2 & 0.583 (0.696) & \textbf{0.781} (0.606) & 0.889 & \textbf{0.825} & \textbf{0.707}\\
\hspace{1em}P1 \& P2 & \textbf{0.656} (0.629) & 0.698 (0.435) & \textbf{0.947} & 0.642 & 0.619\\
\hspace{1em}P1 & 0.819 (0.817) & 0.744 (0.595) & 0.981 & 0.838 & 0.889\\
\hspace{1em}P2 & 0.881 (0.853) & 0.887 (0.942) & 0.926 & 0.719 & 0.776\\

\bottomrule
\end{tabular}
\end{adjustbox}
\end{table*}


\smallskip\noindent\textbf{Worker effort}

\noindent Although our main focus in this paper is quality, we complement our analysis by looking at the impact on worker effort. 
We consider \textit{decision time} as a proxy to estimate the effort incurred on workers.

The median decision time in the baseline condition was $22.66$s for Exergame-VR, 33.88s for AMZ-reviews, $30.77$s for Content-moderation, $23.38$s for Verification, and $25.59$s for Inequality-OA. 
%
In the \textit{80-20} datasets, the median decision time in the baseline condition was $33.62$s for Exergame-VR, and $33.05$s for Inequality-OA.

While formulating complex predicates as simpler multipart questions offer gains in quality, it results in slower task completion time.
Workers in the \pp~condition spent significantly more time than the \textit{baseline} in all datasets ($p < .01$), which intuitively makes sense since workers answered two questions rather than one (the decision time ranged between $36.77$s and $53.56$s).
Likewise, the \pandp~condition was also significantly slower than the \textit{baseline} ($p < .01$, with decision time between 36.96s and 45.08s)\footnote{To approximate the decision time for~\pandp, we determine the median decision time (per document) for conditions $P_1$ and $P_2$ separately. Then for each document, we use the ``slower" predicate as the decision time.}.
%
%
Also, there is no substantial evidence to suggest which task alternative (\pp~or \pandp) is better in terms of effort.
The conditions \pp~and \pandp~had comparable results in 3 out of 5 datasets ($p > .05$), and \pandp~outperformed on the rest ($p < .05$).

Looking closer into the performance, we noticed that simpler predicates (when viewed in isolation) could potentially be faster than asking a complex predicate, but not always. 
When comparing $P_1$ and $P_2$ to the baseline, we noticed two competing observations.
One of the simpler predicates was significantly faster than the baseline in some cases (40\% faster for AMZ-reviews, 27\% for Content-moderation, and 56\% Verification) while significantly slower in some others (20\% slower for AMZ-reviews, 55\% for Verification, and 32\% Inequality-OA).
A similar result can also be observed in the \textit{80-20} scenario.
This suggests worker strategies such as short-circuit evaluation or focusing on simpler criteria when evaluating complex predicates, but the behavior requires further exploration.

To complement our analysis, we also explore how the predicate formulation may have influenced task intake.
Overall, the percentage of workers who quit after a quick inspection of the task (during training) ranged between 18\% and 73\%.
In particular, the task abandonment in the \pp~condition ranged between 50\% and 73\% (somewhat expected given that workers faced the same amount of instructions as in the baseline and had to answer two questions rather than one).
Across all datasets, either $P_1$ or $P_2$ obtained the highest task intake, aligning with the observation from the previous paragraph.
To aid task intake, as task designers, we may seek to formulate a complex predicate as multiple (focused and simpler) questions and query them in isolation. Also,  the instructions length should be kept in mind, in the $P_1$ and $P_2$ conditions our instructions were between 21\% and 55\% shorter than the baseline and \pp. 
However, this suggestion demands further research, and we find it an interesting direction to explore.


Our results show that, as current literature suggests, there is a trade-off between quality and time. Besides, formulating complex predicates as multipart questions could also help identify which predicates may be more effort-intensive.
Please refer to our supplementary material for a more detailed analysis.

\subsection{Simulations}

\noindent The high dimensionality of the problem makes it intractable to crowdsource for every possible parameter value. Here, we rely on simulations to assess how the performance of workers could vary under different parameterizations of the problem.

\textbf{Conditions.} 
The \textit{baseline} task asks the complex predicate $\mathcal{P}$ directly, the \textit{same-task} alternative queries the simpler predicates $\{p_1,\dots,p_n\}$ in one task (i.e., a worker answers $n$ questions), and \textit{separate-tasks} delivers these predicates on different tasks (i.e., a worker answers one of the $p_j$ predicates). 
%
%
We use the terms conditions, cases and task alternatives, interchangeably. 


\textbf{Parameters \& Metric.} We parameterize the simulations based on 1) the number of simpler predicates $n$ that constitute $\mathcal{P}$, 2) the selectivity $s_j$ for predicates $p_j \in \mathcal{P}$, 3) the accuracy of workers drawn from a Beta distribution with mean $\mu$ and variance $\sigma^2$, 4)  the budget $b$ controlling the number of votes per item, and 5) the penalty $\gamma$ that impacts the accuracy of the complex predicate $\mathcal{P}$.
%
To assess different quality goals, we use $F_\beta$, for several values of $\beta$.

\textbf{Worker accuracy}. For a complex $\mathcal{P}$, the \textit{separate-tasks} condition defines a beta distribution for each predicate with expected accuracy $\mu_j$ for $p_j \in \mathcal{P}$. The \textit{same-task} condition defines a beta distribution with accuracy $\mu_s = \frac{1}{n} \sum u_j$. 
In contrast, the \textit{baseline} defines a beta distribution with expected accuracy $\mu_b$ but adjusted based on the penalty $\gamma$.

We describe results for settings without penalty ($\gamma=0$) and summarize the impact of $\gamma$ at the end of this section, referring readers to our supplementary materials for further details on our parameterization and in-depth analysis.

\begin{figure}[h]
  \centering    
  \includegraphics[width=0.9\columnwidth]{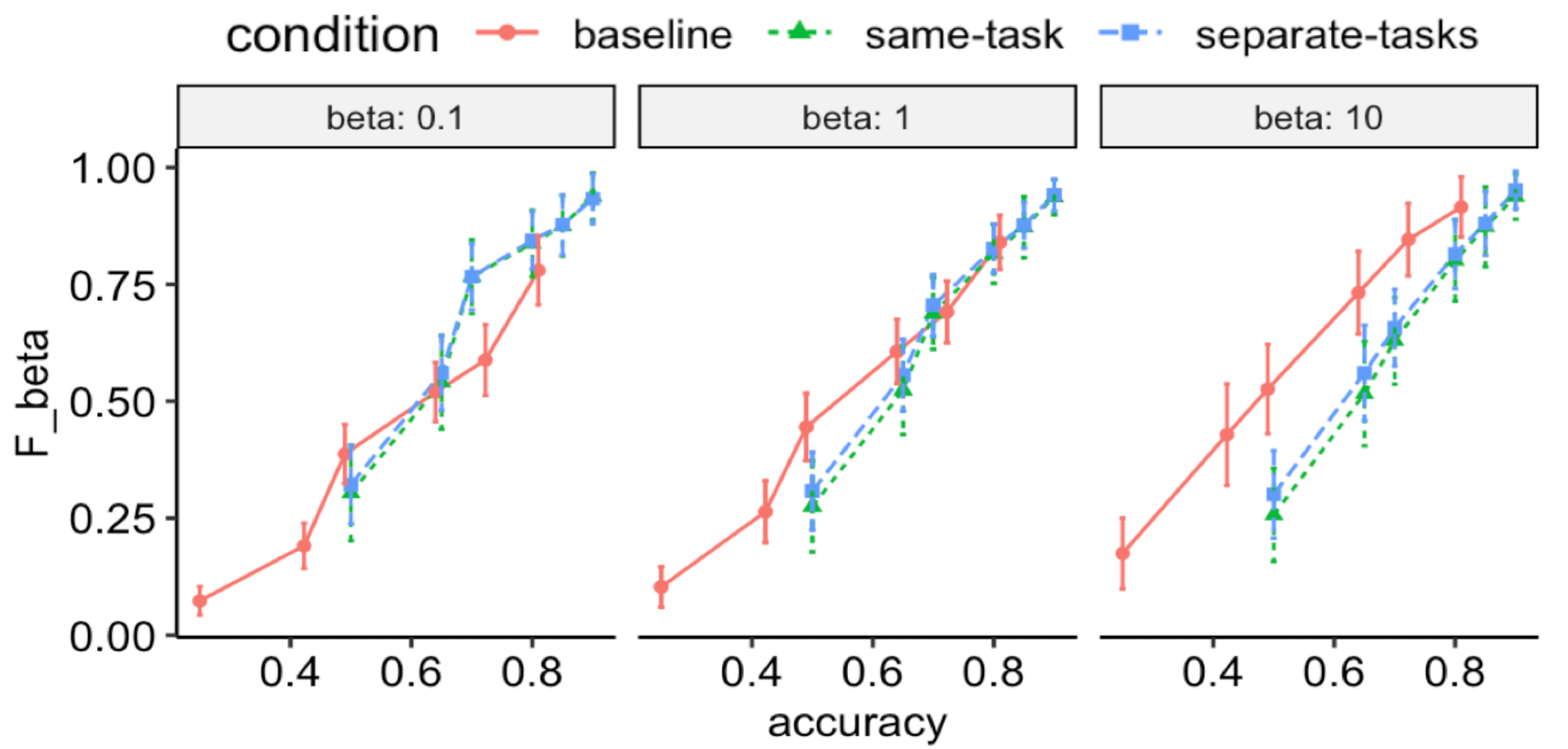}
  \caption{\textmd{Classification performance for different accuracy values, number of predicates $n=2$ and selectivity $s = 0.5$.}}~\label{fig:simulations-1}
\end{figure}

\smallskip\noindent \textbf{Equal selectivity and accuracy}. In this scenario, we define that predicates have equal selectivity $s$, and workers come from the same distribution. Figure \ref{fig:simulations-1} depicts the results for $n=2$, $s=0.5$, and for different expected accuracy values.
It can be noticed than when precision and recall weight equally ($\beta=1$), there is a difference between the task alternatives in favor of the baseline. However, the gap decreases as the accuracy of workers increases (until the conditions perform roughly the same). 
The same-task and separate-tasks alternatives outperform the baseline when precision is more relevant than recall ($\beta=.1$) and workers are better than random ($\mu \ge 0.6$). However, the conditions perform roughly the same when we consider higher selectivity ($s > 0.5$). 
The baseline outperforms the other alternatives when we value more recall ($\beta=10$), and the difference holds as we increase the accuracy and selectivity (except for $s = 0.1$, a extremely low selectivity with high variance).

Increasing the budget (number of votes) does not affect the results in low-accuracy settings. But when accuracy is high ($\mu \ge 0.7)$, the differences narrow until the conditions perform roughly the same. 
%
The number of predicates, however, harms the baseline performance, making the same- and separate-tasks superior choices for all settings.

%
These observations suggest that we may seek to formulate a complex predicate as a single question if we aim to optimize recall and the number of simpler predicates $n$ is low, which intuitively makes sense and aligns with current guidelines for multi-class classification \cite{DBLP:conf/lrec/SabouBDS14}. 
While if we aim for precision or face scenarios with many predicates, we are better off by querying a complex predicate via its constituents.
However, in the following, we assess more realistic settings (different selectivities and accuracies), and see how asking the simpler questions is preferable over $\mathcal{P}$.

\smallskip\noindent \textbf{Different selectivity and same accuracy}. We assign different selectivity values (either low or high) to the predicates, and we assume the same expected accuracy for the individual predicates $p_j$.
We first considered two predicates $p_1$ and $p_2$ and two scenarios where the predicates have selectivities 1) $s_1 = 0.3$ and $s_2=0.7$; and 2) $s_1 = 0.7$ and $s_2 = 0.3$. We tested different accuracies $\mu \in [0.5,0.9]$. 

The results showed the same trend (figure omitted to save space) as in the simulations where we set predicates with equal selectivity and accuracy.
Likewise, varying the number of predicates harmed the baseline performance, and increasing the budget narrowed the difference between the conditions when accuracy is high (until the task alternatives performed roughly equal).

\smallskip\noindent \textbf{Different selectivity and accuracy}. 
Here we consider predicates with different selectivities and accuracies, a setting that aligns better with what we observed in the real crowdsourcing experiment.
First, we simulated two predicates $p_1$ and $p_2$ with selectivity $s_1$ and $s_2$, and expected accuracy $\mu_1$ and $\mu_2$, respectively (where $s_1 \ne s_2$ and $\mu_1 \ne \mu_2$). We considered selectivity values $s_j \in \{0.3, 0.7\}$, a fixed accuracy $\mu_1 \in \{.6, .9\}$ and varied accuracy for $\mu_2$ with $\mu_2 \in [0.6, 0.9]$. We tested all combinations combinations of selectivity and accuracy.

\begin{figure}[h]
  \centering    
  \includegraphics[width=0.9\columnwidth]{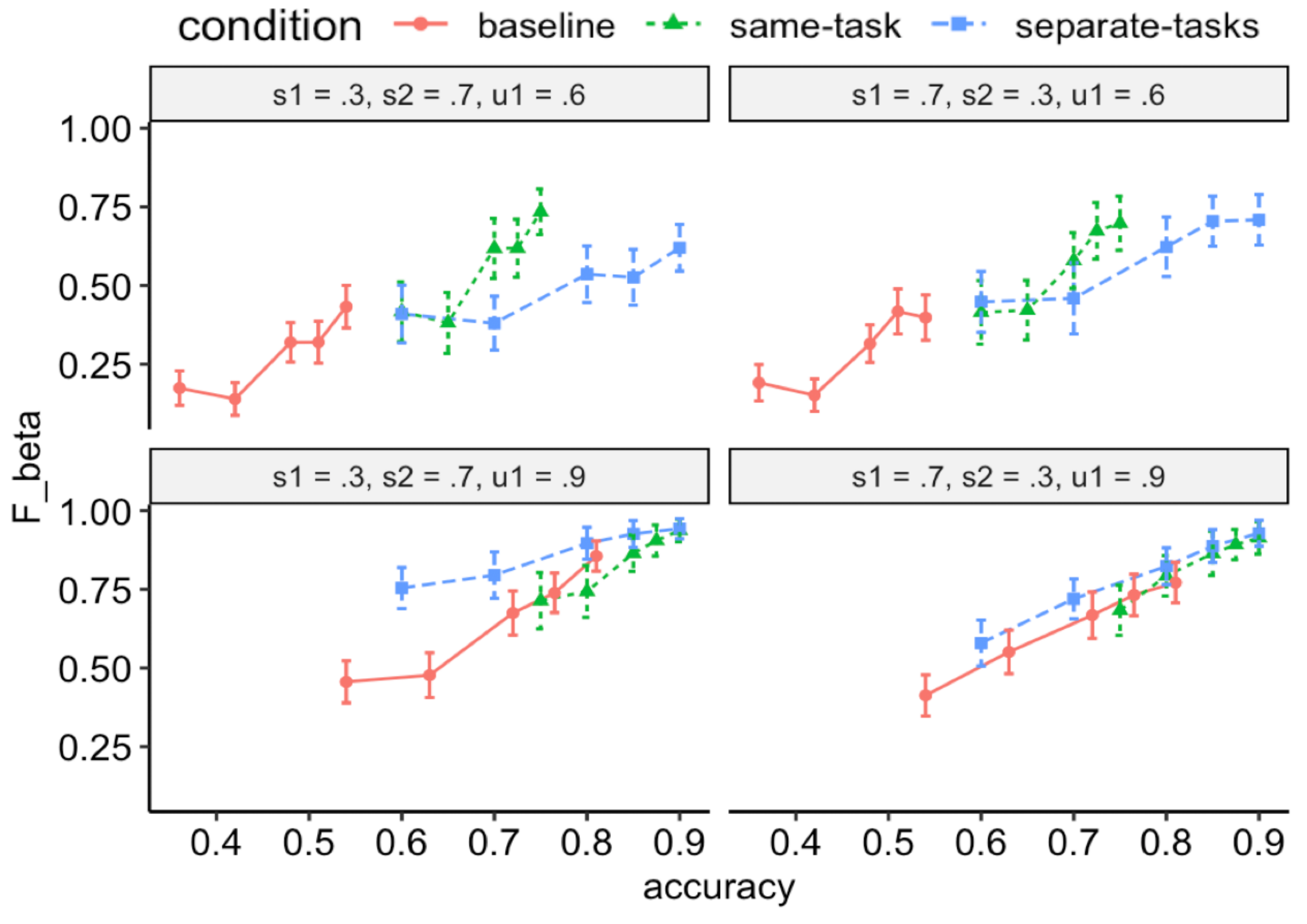}
  \caption{\textmd{Classification performance for predicates with different selectivity and accuracy, $n = 2$ and $\beta = 1$.}}~\label{fig:simulations-2}
\end{figure}

Figure \ref{fig:simulations-2} shows the results for $\beta=1$ (more details in the supplementary material). When we weight recall and precision equally, we noticed a difference in performance in favor of the same- and separate-tasks conditions (though less pronounced for $\mu_1 = .9$).
Like in previous simulations, putting more weight to precision favors the same- and separate-tasks conditions.
As for $\beta \ge 2$, the same- and separate-tasks condition also showed superior performance for the settings where the first predicate had an accuracy $u_1 = 0.6$. In contrast, for high-accuracy settings, $u_2 = 0.9$, the difference between the baseline and separate-task conditions narrowed until these performed roughly the same (both better than same-task).

We also considered the case of multiple predicates ($n = 4$) with different accuracies and selectivities. Like in previous simulations, a higher number of predicates hurts the baseline performance. In this setting, the same- and separate-tasks conditions outperformed the baseline across different values of budget $b$.

\smallskip\noindent\textbf{Summary}. 
Our simulations without penalty showed how formulating a composite predicate as a single question is preferable for recall if we consider a small number of predicates with equal selectivity and accuracy.
However, this is not always the case in real-world settings, where we have many predicates with different accuracy and selectivity. 
%
In these contexts, we noticed that formulating a complex predicate $\mathcal{P}$ as multiple simpler questions showed superior performance in general, which aligns with our real-world experiment.
As we increase the penalty ($\gamma > 0$), the baseline tends towards 0.5 (random guessing), and naturally, the performance deteriorates, making the conditions that ask the individual predicates more suitable.
\section{Hybrid classification}~\label{sec:hybrid}

\subsection{Problem definition}
We extend the crowdsourced classification by allowing to employ a set $M$ of machine learning (ML) classifiers. We want to identify the items in $I$ that meet the complex predicate $\mathcal{P}$, but we now can use ML classifiers alongside crowd workers.

%

To solve this problem, we now consider training ML classifiers as we cast votes from the crowd $\mathcal{W}$. The classifiers $\mathcal{M}$ can be trained for $\mathcal{P}$ directly, or for (some of) the simpler predicates $p_j \in \mathcal{P}$.
Therefore, the solution space is naturally impacted by how well the ML classifiers can learn $\mathcal{P}$ or the individual constituents, and thus help in the crowdsourced classification problem. 





\subsection{Experiment}





\begin{table*}
\caption{\textmd{\label{table:hybrid-summary}
The cells correspond to $F_1$ scores for the best crowd performance (\pp~vs. \pandp), the best ML result on average (among single and ensemble of classifiers), and the best hybrid performance (Crowd-ML vs. ML-Crowd). The standard deviation for ML is $\le 0.07$.}}
\centering
\begin{adjustbox}{width=.8\textwidth}
\fontsize{0.5em}{0.6em}\selectfont
\begin{tabular}[t]{lrrrr}
\toprule
\textbf{Classifier} & \textbf{Exergame-VR} & \textbf{Inequality-OA} & \textbf{AMZ-reviews} & \textbf{Content-moderation} \\

 \textit{Distribution} & \textit{60-40 (80-20)} & \textit{60-40 (80-20)} & \textit{60-40}& \textit{60-40}\\
 \midrule

\hspace{1em}Crowd & 0.656 (0.696) & 0.781 (0.606) & \textbf{0.947} & \textbf{0.825}\\
\hspace{1em}ML & \textbf{0.866} (0.821) & \textbf{0.853} (0.651) & 0.753 & 0.183\\
\hspace{1em}Hybrid & 0.775 (0.762) & 0.800 (0.588) & 0.931 & 0.485\\

\bottomrule
\end{tabular}
\end{adjustbox}
\end{table*}

The crowdsourcing experiment showed how performance gains are obtained by querying a complex predicate as multiple (simpler) questions and then combining back the results.
Here we situate the predicate formulation problem in the context of hybrid classification and test our insight from the crowdsourcing experiment.
%
The literature suggests that hybrid classification offers superior results. And our intuition is that formulating a complex predicate as multiple questions would allow us to capitalize on the strength of crowd and ML classifiers and, therefore, obtain superior performance.




\textbf{Design}. We consider four datasets from the crowdsourcing experiment (excluding \textit{Verification}), with 118 items in each dataset.
The crowd judgments from the crowdsourcing experiment are aggregated using majority voting, and we combine these with machine predictions in two (simplistic) ways: Crowd-ML and ML-Crowd, where \textit{Crowd-ML} leverages the crowd for the first predicate ($p_1$) and machine for the second ($p_2$), while \textit{ML-Crowd} does the opposite (computing $p_1 \land p_2$ to derive the complex $\mathcal{P}$).

We use classifiers and ensembles of classifiers in this experiment. The four machine learning classifiers correspond to Logistic Regression (LR), Support Vector Machine (SVM), BERT \cite{devlin2018bert} and DistilBERT \cite{DistilBERT2019}.
%
The aim of covering different ML techniques is to give our analysis breadth and not to compare the models, primarily since we are operating with small datasets.
The ensemble methods use LR, SVM, and Multinomial Naive Bayes (MNB) as base estimators. We considered voting classifiers (``hard", using majority voting, and ``soft", using the predicted probabilities), a bagging classifier (with SVM as its base estimator), and a stacking classifier.

The models were trained on the complex $\mathcal{P}$ and its constituents.
We used 10-fold stratified cross-validation, repeating the experiment 10 times (with different seeds) and reporting averages. We fine-tuned the deep learning models for $4$ epochs with a learning rate of $0.001$ using the AdamW optimizer \cite{DBLP:conf/iclr/LoshchilovH19}.
%
We used an over-sampling technique \cite{SMOTE2002} to aid the LR, SVM, and MNB classifiers (alone and within an ensemble) in dealing with imbalanced classes.
%

\subsection{Results}

Hybrid classification, Table \ref{table:hybrid-summary}, showed a superior (or comparable) performance when compared to crowd classification for most of the datasets we considered (see our supplementary material for a more in-depth analysis).
For the \textit{60-40} case, the hybrid classifier outperformed the crowd for the Exergame-VR and Inequality-OA datasets ($16\%$ and $2\%$ difference in performance, respectively). For AMZ-reviews, the performance was comparable (both classifiers with $F_1 > 0.9$) while for Content-moderation the crowd showed superior classification with a score of $F_1=0.82$ in comparison to only $0.48$ for the hybrid approach (this was a difficult dataset in general for both crowd and ML classifiers). 
The hybrid classification outperformed in the \textit{80-20} variant of the Exergame-VR dataset ($9\%$ difference in $F_1$), while the crowd obtained a slightly better performance for the \textit{80-20} version of Inequality-OA ($3\%$ difference).

Hybrid classification outperformed ML for AMZ-reviews (21\% difference) and Content-moderation datasets, although the hybrid performance was almost random for Content-moderation. 
%
%
In contrast, ML performed better for Exergame-VR and Inequality-OA datasets (11\% and 6\% difference, respectively), including the imbalanced variants, where the difference was at most 10\%.

From a task design perspective, these results suggest that framing a complex predicate as multiple simpler questions translates into performance gains and plays nicely with recommendations from hybrid classification research.
%
Querying a complex predicate $\mathcal{P}$ via its constituents allows for a (potentially) better coupling of crowd and machine classifiers. 
Our experiment showed that even this simple Human-AI collaboration approach gives a solid and consistent performance across different datasets and domains.



\section{Discussion \& Conclusion}~\label{sec:discussion}

Performance gains could be obtained depending on how we formulate a composite question in the context of crowdsourced and hybrid classification.
From a task designer perspective, leveraging focused more straightforward questions offers more detailed information about crowd workers, and \textit{can inform the use of different approaches more adapted to the characteristics (difficulty, selectivity) of each simpler predicate}, instead of committing to a single strategy (e.g., hiring different workers based on task difficulty  \cite{DBLP:journals/pvldb/HaasAGM15,DBLP:journals/pacmhci/RetelnyBV17}).
%

Querying simpler predicates could enable \textit{more effective coupling of ML classifiers and favor long term reusability of already trained models}.
We believe that there is potential for training highly-specialized models that couple effectively with the performance of workers (instead of learning models classify items based on complex predicates directly).
Besides, answering simpler questions outputs reusable (and detailed) knowledge about the capabilities of crowd and machine classifiers. 
For example, if we were to work on an SLR about \textit{exergame usage in older adults}, we could rely on the current knowledge that we have built by querying the simpler predicates from the Exergame-VR and Inequality-OA datasets. From the perspective of crowd workers, this means reapplying learned skills, and for machines, it involves classifying unseen papers (and filter out at least articles that are ``obviously" not relevant).

We focused on a specific but relevant aspect for task designers: how to frame a composite question used to classify items.
Our results 
showed that superior classification performance could be obtained by querying a complex predicate as multiple (simpler) questions instead of asking a single coarse predicate.
In a scenario with low accuracy and selectivity, asking the constituents of $\mathcal{P}$ (i.e., $n$ questions) may increase the chances of misclassifying items, as observed in our simulations. 
In this case, we may rely on framing the complex $\mathcal{P}$ as a single question (limited by the number of predicates it contains) or framing $\mathcal{P}$ as a mix of simpler and coarse questions. 
To some extent, our competing results from either asking predicates on the same task vs. on separate tasks is related to this point (i.e., the error rate of a single worker answering $n$ questions vs. $n$ workers answering a question each). 
%
%
Both task design choices offer superior results over the baseline, but there is not enough evidence to inform decisions based on given problem settings.
We find this an interesting direction of future work, where we design algorithms that model workers, tasks, and predicates to automatically learn how to formulate complex predicates to meet quality goals while operating under a budget.

\begin{acks}
This work was supported by the Russian Science Foundation (Project No. 19-18-00282).
\end{acks}

\bibliographystyle{ACM-Reference-Format}
\bibliography{shorter-references}


\begin{thebibliography}{53}


\ifx \showCODEN    \undefined \def \showCODEN     #1{\unskip}     \fi
\ifx \showDOI      \undefined \def \showDOI       #1{#1}\fi
\ifx \showISBNx    \undefined \def \showISBNx     #1{\unskip}     \fi
\ifx \showISBNxiii \undefined \def \showISBNxiii  #1{\unskip}     \fi
\ifx \showISSN     \undefined \def \showISSN      #1{\unskip}     \fi
\ifx \showLCCN     \undefined \def \showLCCN      #1{\unskip}     \fi
\ifx \shownote     \undefined \def \shownote      #1{#1}          \fi
\ifx \showarticletitle \undefined \def \showarticletitle #1{#1}   \fi
\ifx \showURL      \undefined \def \showURL       {\relax}        \fi
\providecommand\bibfield[2]{#2}
\providecommand\bibinfo[2]{#2}
\providecommand\natexlab[1]{#1}
\providecommand\showeprint[2][]{arXiv:#2}

\bibitem[\protect\citeauthoryear{Chandler and Kapelner}{Chandler and
  Kapelner}{2013}]%
        {CHANDLER2013Breaking}
\bibfield{author}{\bibinfo{person}{Dana Chandler} {and} \bibinfo{person}{Adam
  Kapelner}.} \bibinfo{year}{2013}\natexlab{}.
\newblock \showarticletitle{Breaking monotony with meaning: Motivation in
  crowdsourcing markets}.
\newblock \bibinfo{journal}{\emph{Journal of Economic Behavior \&
  Organization}}  \bibinfo{volume}{90} (\bibinfo{year}{2013}),
  \bibinfo{pages}{123--133}.
\newblock
\showISSN{0167-2681}


\bibitem[\protect\citeauthoryear{Cheng and Bernstein}{Cheng and
  Bernstein}{2015}]%
        {DBLP:conf/cscw/ChengB15}
\bibfield{author}{\bibinfo{person}{Justin Cheng} {and}
  \bibinfo{person}{Michael~S. Bernstein}.} \bibinfo{year}{2015}\natexlab{}.
\newblock \showarticletitle{Flock: Hybrid Crowd-Machine Learning Classifiers}.
  In \bibinfo{booktitle}{\emph{{CSCW} 2015}}.
\newblock


\bibitem[\protect\citeauthoryear{Devlin}{Devlin}{2018}]%
        {devlin2018bert}
\bibfield{author}{\bibinfo{person}{Jacob et~al. Devlin}.}
  \bibinfo{year}{2018}\natexlab{}.
\newblock \showarticletitle{BERT: Pre-training of Deep Bidirectional
  Transformers for Language Understanding}.
\newblock \bibinfo{journal}{\emph{arXiv preprint arXiv:1810.04805}}
  (\bibinfo{year}{2018}).
\newblock


\bibitem[\protect\citeauthoryear{Dunn}{Dunn}{1964}]%
        {DunnTest1964}
\bibfield{author}{\bibinfo{person}{Olive~Jean Dunn}.}
  \bibinfo{year}{1964}\natexlab{}.
\newblock \showarticletitle{Multiple Comparisons Using Rank Sums}.
\newblock \bibinfo{journal}{\emph{Technometrics}} \bibinfo{volume}{6},
  \bibinfo{number}{3} (\bibinfo{year}{1964}).
\newblock
\showISSN{00401706}


\bibitem[\protect\citeauthoryear{et~al.}{et~al.}{2012a}]%
        {DBLP:conf/sigmod/ParameswaranGPPRW12}
\bibfield{author}{\bibinfo{person}{Aditya G.~Parameswaran et al.}}
  \bibinfo{year}{2012}\natexlab{a}.
\newblock \showarticletitle{CrowdScreen: algorithms for filtering data with
  humans}. In \bibinfo{booktitle}{\emph{{SIGMOD} 2012}}.
\newblock


\bibitem[\protect\citeauthoryear{et~al.}{et~al.}{2017a}]%
        {DBLP:journals/pvldb/JainSPW17}
\bibfield{author}{\bibinfo{person}{Ayush~Jain et al.}}
  \bibinfo{year}{2017}\natexlab{a}.
\newblock \showarticletitle{Understanding Workers, Developing Effective Tasks,
  and Enhancing Marketplace Dynamics: {A} Study of a Large Crowdsourcing
  Marketplace}.
\newblock \bibinfo{journal}{\emph{{PVLDB}}} \bibinfo{volume}{10},
  \bibinfo{number}{7} (\bibinfo{year}{2017}).
\newblock


\bibitem[\protect\citeauthoryear{et~al.}{et~al.}{2016}]%
        {DBLP:conf/naacl/LiuSBLLW16}
\bibfield{author}{\bibinfo{person}{Angli~Liu et al.}}
  \bibinfo{year}{2016}\natexlab{}.
\newblock \showarticletitle{Effective Crowd Annotation for Relation
  Extraction}. In \bibinfo{booktitle}{\emph{{NAACL} {HLT} 2016}}.
\newblock


\bibitem[\protect\citeauthoryear{et~al.}{et~al.}{2015a}]%
        {DBLP:conf/www/HoSSV15}
\bibfield{author}{\bibinfo{person}{Chien{-}Ju~Ho et al.}}
  \bibinfo{year}{2015}\natexlab{a}.
\newblock \showarticletitle{Incentivizing High Quality Crowdwork}. In
  \bibinfo{booktitle}{\emph{{WWW} 2015}}.
\newblock


\bibitem[\protect\citeauthoryear{et~al.}{et~al.}{2018a}]%
        {DBLP:conf/wsdm/DifallahFI18}
\bibfield{author}{\bibinfo{person}{Djellel Eddine~Difallah et al.}}
  \bibinfo{year}{2018}\natexlab{a}.
\newblock \showarticletitle{Demographics and Dynamics of Mechanical Turk
  Workers}. In \bibinfo{booktitle}{\emph{{WSDM} 2018}}.
  \bibinfo{pages}{135--143}.
\newblock


\bibitem[\protect\citeauthoryear{et~al.}{et~al.}{2017b}]%
        {DBLP:conf/hcomp/LanRST17}
\bibfield{author}{\bibinfo{person}{Doren~Lan et al.}}
  \bibinfo{year}{2017}\natexlab{b}.
\newblock \showarticletitle{Dynamic Filter: Adaptive Query Processing with the
  Crowd}. In \bibinfo{booktitle}{\emph{{HCOMP} 2017}}.
\newblock


\bibitem[\protect\citeauthoryear{et~al.}{et~al.}{2017c}]%
        {KrivosheevHcomp2017}
\bibfield{author}{\bibinfo{person}{Evgeny~Krivosheev et al.}}
  \bibinfo{year}{2017}\natexlab{c}.
\newblock \showarticletitle{Crowdsourcing Paper Screening in Systematic
  Literature Reviews}. In \bibinfo{booktitle}{\emph{{HCOMP} 2017}}.
\newblock


\bibitem[\protect\citeauthoryear{et~al.}{et~al.}{2018b}]%
        {DBLP:journals/csur/DanielKCBA18}
\bibfield{author}{\bibinfo{person}{Florian~Daniel et al.}}
  \bibinfo{year}{2018}\natexlab{b}.
\newblock \showarticletitle{Quality Control in Crowdsourcing: {A} Survey of
  Quality Attributes, Assessment Techniques, and Assurance Actions}.
\newblock \bibinfo{journal}{\emph{{ACM} Comput. Surv.}} \bibinfo{volume}{51},
  \bibinfo{number}{1} (\bibinfo{year}{2018}), \bibinfo{pages}{7:1--7:40}.
\newblock
\urldef\tempurl%
\url{https://doi.org/10.1145/3148148}
\showDOI{\tempurl}


\bibitem[\protect\citeauthoryear{et~al.}{et~al.}{2010}]%
        {DBLP:conf/uist/LittleCGM10}
\bibfield{author}{\bibinfo{person}{Greg~Little et al.}}
  \bibinfo{year}{2010}\natexlab{}.
\newblock \showarticletitle{TurKit: human computation algorithms on mechanical
  turk}. In \bibinfo{booktitle}{\emph{{UIST} 2010}}.
\newblock


\bibitem[\protect\citeauthoryear{et~al.}{et~al.}{2015b}]%
        {DBLP:conf/chi/ChengTIB15}
\bibfield{author}{\bibinfo{person}{Justin~Cheng et al.}}
  \bibinfo{year}{2015}\natexlab{b}.
\newblock \showarticletitle{Break It Down: {A} Comparison of Macro- and
  Microtasks}. In \bibinfo{booktitle}{\emph{{CHI} 2015}}.
  \bibinfo{pages}{4061--4064}.
\newblock
\urldef\tempurl%
\url{https://doi.org/10.1145/2702123.2702146}
\showDOI{\tempurl}


\bibitem[\protect\citeauthoryear{et~al.}{et~al.}{2019}]%
        {DBLP:conf/wsdm/HanRGSCMD19}
\bibfield{author}{\bibinfo{person}{Lei~Han et al.}}
  \bibinfo{year}{2019}\natexlab{}.
\newblock \showarticletitle{All Those Wasted Hours: On Task Abandonment in
  Crowdsourcing}. In \bibinfo{booktitle}{\emph{{WSDM} 2019}}.
\newblock
\urldef\tempurl%
\url{https://doi.org/10.1145/3289600.3291035}
\showDOI{\tempurl}


\bibitem[\protect\citeauthoryear{et~al.}{et~al.}{2011}]%
        {DBLP:conf/sigmod/FranklinKKRX11}
\bibfield{author}{\bibinfo{person}{Michael J.~Franklin et al.}}
  \bibinfo{year}{2011}\natexlab{}.
\newblock \showarticletitle{CrowdDB: answering queries with crowdsourcing}. In
  \bibinfo{booktitle}{\emph{{SIGMOD} 2011}}.
\newblock


\bibitem[\protect\citeauthoryear{et~al.}{et~al.}{2014}]%
        {DBLP:conf/lrec/SabouBDS14}
\bibfield{author}{\bibinfo{person}{Marta~Sabou et al.}}
  \bibinfo{year}{2014}\natexlab{}.
\newblock \showarticletitle{Corpus Annotation through Crowdsourcing: Towards
  Best Practice Guidelines}. In \bibinfo{booktitle}{\emph{{LREC} 2014}}.
  \bibinfo{pages}{859--866}.
\newblock


\bibitem[\protect\citeauthoryear{et~al.}{et~al.}{2002}]%
        {SMOTE2002}
\bibfield{author}{\bibinfo{person}{Nitesh V.~Chawla et al.}}
  \bibinfo{year}{2002}\natexlab{}.
\newblock \showarticletitle{{SMOTE:} Synthetic Minority Over-sampling
  Technique}.
\newblock \bibinfo{journal}{\emph{J. Artif. Intell. Res.}}
  \bibinfo{volume}{16} (\bibinfo{year}{2002}), \bibinfo{pages}{321--357}.
\newblock
\urldef\tempurl%
\url{https://doi.org/10.1613/jair.953}
\showDOI{\tempurl}


\bibitem[\protect\citeauthoryear{et~al.}{et~al.}{2012b}]%
        {DBLP:conf/cscw/DowKKH12}
\bibfield{author}{\bibinfo{person}{Steven~Dow et al.}}
  \bibinfo{year}{2012}\natexlab{b}.
\newblock \showarticletitle{Shepherding the crowd yields better work}. In
  \bibinfo{booktitle}{\emph{{CSCW} 2012}}. \bibinfo{pages}{1013--1022}.
\newblock
\urldef\tempurl%
\url{https://doi.org/10.1145/2145204.2145355}
\showDOI{\tempurl}


\bibitem[\protect\citeauthoryear{et~al.}{et~al.}{2018c}]%
        {DBLP:journals/pacmhci/CallaghanGMLL18}
\bibfield{author}{\bibinfo{person}{William~Callaghan et al.}}
  \bibinfo{year}{2018}\natexlab{c}.
\newblock \showarticletitle{MechanicalHeart: {A} Human-Machine Framework for
  the Classification of Phonocardiograms}.
\newblock \bibinfo{journal}{\emph{{PACMHCI}}} \bibinfo{volume}{2},
  \bibinfo{number}{{CSCW}} (\bibinfo{year}{2018}).
\newblock


\bibitem[\protect\citeauthoryear{Faltings, Jurca, Pu, and Tran}{Faltings
  et~al\mbox{.}}{2014}]%
        {DBLP:conf/hcomp/FaltingsJPT14}
\bibfield{author}{\bibinfo{person}{Boi Faltings}, \bibinfo{person}{Radu Jurca},
  \bibinfo{person}{Pearl Pu}, {and} \bibinfo{person}{Bao~Duy Tran}.}
  \bibinfo{year}{2014}\natexlab{}.
\newblock \showarticletitle{Incentives to Counter Bias in Human Computation}.
  In \bibinfo{booktitle}{\emph{{HCOMP} 2014}}.
\newblock


\bibitem[\protect\citeauthoryear{Gadiraju and Kawase}{Gadiraju and
  Kawase}{2017}]%
        {gadiraju2017improving}
\bibfield{author}{\bibinfo{person}{Ujwal Gadiraju} {and}
  \bibinfo{person}{Ricardo Kawase}.} \bibinfo{year}{2017}\natexlab{}.
\newblock \showarticletitle{Improving reliability of crowdsourced results by
  detecting crowd workers with multiple identities}. In
  \bibinfo{booktitle}{\emph{{ICWE} 2017}}. Springer, \bibinfo{pages}{190--205}.
\newblock


\bibitem[\protect\citeauthoryear{Gadiraju, Kawase, and Dietze}{Gadiraju
  et~al\mbox{.}}{2014}]%
        {DBLP:conf/ht/GadirajuKD14}
\bibfield{author}{\bibinfo{person}{Ujwal Gadiraju}, \bibinfo{person}{Ricardo
  Kawase}, {and} \bibinfo{person}{Stefan Dietze}.}
  \bibinfo{year}{2014}\natexlab{}.
\newblock \showarticletitle{A taxonomy of microtasks on the web}. In
  \bibinfo{booktitle}{\emph{{HT} 2014}}. \bibinfo{pages}{218--223}.
\newblock
\urldef\tempurl%
\url{https://doi.org/10.1145/2631775.2631819}
\showDOI{\tempurl}


\bibitem[\protect\citeauthoryear{Gadiraju, Yang, and Bozzon}{Gadiraju
  et~al\mbox{.}}{2017}]%
        {DBLP:conf/ht/GadirajuYB17}
\bibfield{author}{\bibinfo{person}{Ujwal Gadiraju}, \bibinfo{person}{Jie Yang},
  {and} \bibinfo{person}{Alessandro Bozzon}.} \bibinfo{year}{2017}\natexlab{}.
\newblock \showarticletitle{Clarity is a Worthwhile Quality: On the Role of
  Task Clarity in Microtask Crowdsourcing}. In \bibinfo{booktitle}{\emph{{HT}
  2017}}.
\newblock


\bibitem[\protect\citeauthoryear{Haas, Ansel, Gu, and Marcus}{Haas
  et~al\mbox{.}}{2015}]%
        {DBLP:journals/pvldb/HaasAGM15}
\bibfield{author}{\bibinfo{person}{Daniel Haas}, \bibinfo{person}{Jason Ansel},
  \bibinfo{person}{Lydia Gu}, {and} \bibinfo{person}{Adam Marcus}.}
  \bibinfo{year}{2015}\natexlab{}.
\newblock \showarticletitle{Argonaut: Macrotask Crowdsourcing for Complex Data
  Processing}.
\newblock \bibinfo{journal}{\emph{{PVLDB}}} \bibinfo{volume}{8},
  \bibinfo{number}{12} (\bibinfo{year}{2015}), \bibinfo{pages}{1642--1653}.
\newblock


\bibitem[\protect\citeauthoryear{Kittur, Chi, and Suh}{Kittur
  et~al\mbox{.}}{2008}]%
        {DBLP:conf/chi/KitturCS08}
\bibfield{author}{\bibinfo{person}{Aniket Kittur}, \bibinfo{person}{Ed~H. Chi},
  {and} \bibinfo{person}{Bongwon Suh}.} \bibinfo{year}{2008}\natexlab{}.
\newblock \showarticletitle{Crowdsourcing user studies with Mechanical Turk}.
  In \bibinfo{booktitle}{\emph{{CHI} 2008}}.
\newblock
\urldef\tempurl%
\url{https://doi.org/10.1145/1357054.1357127}
\showDOI{\tempurl}


\bibitem[\protect\citeauthoryear{Kittur, Nickerson, and et~al.}{Kittur
  et~al\mbox{.}}{2013}]%
        {DBLP:conf/cscw/KitturNBGSZLH13}
\bibfield{author}{\bibinfo{person}{Aniket Kittur}, \bibinfo{person}{Jeffrey~V.
  Nickerson}, {and} \bibinfo{person}{Michael S.~Bernstein et al.}}
  \bibinfo{year}{2013}\natexlab{}.
\newblock \showarticletitle{The future of crowd work}. In
  \bibinfo{booktitle}{\emph{{CSCW} 2013}}.
\newblock
\urldef\tempurl%
\url{https://doi.org/10.1145/2441776.2441923}
\showURL{%
\tempurl}


\bibitem[\protect\citeauthoryear{Krause and Kizilcec}{Krause and
  Kizilcec}{2015}]%
        {DBLP:conf/hcomp/KrauseK15}
\bibfield{author}{\bibinfo{person}{Markus Krause} {and}
  \bibinfo{person}{Ren{\'{e}}~F. Kizilcec}.} \bibinfo{year}{2015}\natexlab{}.
\newblock \showarticletitle{To Play or Not to Play: Interactions between
  Response Quality and Task Complexity in Games and Paid Crowdsourcing}. In
  \bibinfo{booktitle}{\emph{{HCOMP} 2015}}. \bibinfo{pages}{102--109}.
\newblock


\bibitem[\protect\citeauthoryear{Krivosheev~et al.}{Krivosheev~et al.}{2018}]%
        {KrivosheevCSCW2018}
\bibfield{author}{\bibinfo{person}{Evgeny Krivosheev~et al.}}
  \bibinfo{year}{2018}\natexlab{}.
\newblock \showarticletitle{Combining Crowd and Machines for Multi-predicate
  Item Screening}.
\newblock \bibinfo{journal}{\emph{{PACMHCI}}} \bibinfo{volume}{2},
  \bibinfo{number}{CSCW}, Article \bibinfo{articleno}{97} (\bibinfo{date}{Nov.}
  \bibinfo{year}{2018}).
\newblock
\showISSN{2573-0142}


\bibitem[\protect\citeauthoryear{Loshchilov and Hutter}{Loshchilov and
  Hutter}{2019}]%
        {DBLP:conf/iclr/LoshchilovH19}
\bibfield{author}{\bibinfo{person}{Ilya Loshchilov} {and}
  \bibinfo{person}{Frank Hutter}.} \bibinfo{year}{2019}\natexlab{}.
\newblock \showarticletitle{Decoupled Weight Decay Regularization}. In
  \bibinfo{booktitle}{\emph{{ICLR} 2019}}.
\newblock
\urldef\tempurl%
\url{https://openreview.net/forum?id=Bkg6RiCqY7}
\showURL{%
\tempurl}


\bibitem[\protect\citeauthoryear{Mitra, Hutto, and Gilbert}{Mitra
  et~al\mbox{.}}{2015}]%
        {DBLP:conf/chi/MitraHG15}
\bibfield{author}{\bibinfo{person}{Tanushree Mitra},
  \bibinfo{person}{Clayton~J. Hutto}, {and} \bibinfo{person}{Eric Gilbert}.}
  \bibinfo{year}{2015}\natexlab{}.
\newblock \showarticletitle{Comparing Person- and Process-centric Strategies
  for Obtaining Quality Data on Amazon Mechanical Turk}. In
  \bibinfo{booktitle}{\emph{{CHI} 2015}}. \bibinfo{pages}{1345--1354}.
\newblock
\urldef\tempurl%
\url{https://doi.org/10.1145/2702123.2702553}
\showDOI{\tempurl}


\bibitem[\protect\citeauthoryear{Mortensen~et al.}{Mortensen~et al.}{2016}]%
        {Mortensen2016crowd}
\bibfield{author}{\bibinfo{person}{Michael~L. Mortensen~et al.}}
  \bibinfo{year}{2016}\natexlab{}.
\newblock \showarticletitle{An exploration of crowdsourcing citation screening
  for systematic reviews}.
\newblock \bibinfo{journal}{\emph{Research Synthesis Methods}}
  (\bibinfo{year}{2016}).
\newblock
\showISSN{1759-2887}


\bibitem[\protect\citeauthoryear{Nguyen, Wallace, and Lease}{Nguyen
  et~al\mbox{.}}{2015}]%
        {DBLP:conf/hcomp/NguyenWL15}
\bibfield{author}{\bibinfo{person}{An~Thanh Nguyen}, \bibinfo{person}{Byron~C.
  Wallace}, {and} \bibinfo{person}{Matthew Lease}.}
  \bibinfo{year}{2015}\natexlab{}.
\newblock \showarticletitle{Combining Crowd and Expert Labels Using Decision
  Theoretic Active Learning}. In \bibinfo{booktitle}{\emph{{HCOMP} 2015}}.
  \bibinfo{pages}{120--129}.
\newblock
\urldef\tempurl%
\url{http://www.aaai.org/ocs/index.php/HCOMP/HCOMP15/paper/view/11567}
\showURL{%
\tempurl}


\bibitem[\protect\citeauthoryear{Park and Widom}{Park and Widom}{2013}]%
        {DBLP:journals/pvldb/ParkW13}
\bibfield{author}{\bibinfo{person}{Hyunjung Park} {and}
  \bibinfo{person}{Jennifer Widom}.} \bibinfo{year}{2013}\natexlab{}.
\newblock \showarticletitle{Query Optimization over Crowdsourced Data}.
\newblock \bibinfo{journal}{\emph{{PVLDB}}} \bibinfo{volume}{6},
  \bibinfo{number}{10} (\bibinfo{year}{2013}), \bibinfo{pages}{781--792}.
\newblock
\urldef\tempurl%
\url{https://doi.org/10.14778/2536206.2536207}
\showDOI{\tempurl}


\bibitem[\protect\citeauthoryear{Porter, Verdery, and Gaddis}{Porter
  et~al\mbox{.}}{2020}]%
        {Porter2020}
\bibfield{author}{\bibinfo{person}{Nathaniel~D. Porter},
  \bibinfo{person}{Ashton~M. Verdery}, {and} \bibinfo{person}{S.~Michael
  Gaddis}.} \bibinfo{year}{2020}\natexlab{}.
\newblock \showarticletitle{Enhancing big data in the social sciences with
  crowdsourcing: Data augmentation practices, techniques, and opportunities}.
\newblock \bibinfo{journal}{\emph{PLOS ONE}} \bibinfo{volume}{15},
  \bibinfo{number}{6} (\bibinfo{date}{06} \bibinfo{year}{2020}),
  \bibinfo{pages}{1--21}.
\newblock
\urldef\tempurl%
\url{https://doi.org/10.1371/journal.pone.0233154}
\showDOI{\tempurl}


\bibitem[\protect\citeauthoryear{Qarout, Checco, Demartini, and
  Bontcheva}{Qarout et~al\mbox{.}}{2019}]%
        {Qarout2019PlatformRelatedFI}
\bibfield{author}{\bibinfo{person}{Rehab~Kamal Qarout},
  \bibinfo{person}{Alessandro Checco}, \bibinfo{person}{Gianluca Demartini},
  {and} \bibinfo{person}{Kalina Bontcheva}.} \bibinfo{year}{2019}\natexlab{}.
\newblock \showarticletitle{Platform-Related Factors in Repeatability and
  Reproducibility of Crowdsourcing Tasks}. In \bibinfo{booktitle}{\emph{{HCOMP}
  2019}}.
\newblock


\bibitem[\protect\citeauthoryear{Ram{\'i}rez, Baez, Casati, and
  Benatallah}{Ram{\'i}rez et~al\mbox{.}}{2019}]%
        {RamirezBMC2019}
\bibfield{author}{\bibinfo{person}{Jorge Ram{\'i}rez}, \bibinfo{person}{Marcos
  Baez}, \bibinfo{person}{Fabio Casati}, {and} \bibinfo{person}{Boualem
  Benatallah}.} \bibinfo{year}{2019}\natexlab{}.
\newblock \showarticletitle{Crowdsourced dataset to study the generation and
  impact of text highlighting in classification tasks}.
\newblock \bibinfo{journal}{\emph{BMC Research Notes}} \bibinfo{volume}{12},
  \bibinfo{number}{1} (\bibinfo{year}{2019}), \bibinfo{pages}{820}.
\newblock
\showISSN{1756-0500}


\bibitem[\protect\citeauthoryear{Ram\'{i}rez, Baez, Casati, and
  Benatallah}{Ram\'{i}rez et~al\mbox{.}}{2019}]%
        {ramirez2019}
\bibfield{author}{\bibinfo{person}{Jorge Ram\'{i}rez}, \bibinfo{person}{Marcos
  Baez}, \bibinfo{person}{Fabio Casati}, {and} \bibinfo{person}{Boualem
  Benatallah}.} \bibinfo{year}{2019}\natexlab{}.
\newblock \showarticletitle{Understanding the Impact of Text Highlighting in
  Crowdsourcing Tasks}. In \bibinfo{booktitle}{\emph{{HCOMP} 2019}},
  Vol.~\bibinfo{volume}{7}. \bibinfo{publisher}{AAAI},
  \bibinfo{pages}{144--152}.
\newblock


\bibitem[\protect\citeauthoryear{Ram{\'i}rez, Baez, Casati, Cernuzzi, and
  Benatallah}{Ram{\'i}rez et~al\mbox{.}}{2020}]%
        {Ramirez2020DrecCSCW}
\bibfield{author}{\bibinfo{person}{Jorge Ram{\'i}rez}, \bibinfo{person}{Marcos
  Baez}, \bibinfo{person}{Fabio Casati}, \bibinfo{person}{Luca Cernuzzi}, {and}
  \bibinfo{person}{Boualem Benatallah}.} \bibinfo{year}{2020}\natexlab{}.
\newblock \showarticletitle{DREC: towards a Datasheet for Reporting Experiments
  in Crowdsourcing}. In \bibinfo{booktitle}{\emph{CSCW 2020}}.
\newblock


\bibitem[\protect\citeauthoryear{Ram{\'{\i}}rez, Degiacomi, Zanella,
  B{\'{a}}ez, Casati, and Benatallah}{Ram{\'{\i}}rez et~al\mbox{.}}{2019}]%
        {CrowdHub2019}
\bibfield{author}{\bibinfo{person}{Jorge Ram{\'{\i}}rez},
  \bibinfo{person}{Simone Degiacomi}, \bibinfo{person}{Davide Zanella},
  \bibinfo{person}{Marcos B{\'{a}}ez}, \bibinfo{person}{Fabio Casati}, {and}
  \bibinfo{person}{Boualem Benatallah}.} \bibinfo{year}{2019}\natexlab{}.
\newblock \showarticletitle{CrowdHub: Extending crowdsourcing platforms for the
  controlled evaluation of tasks designs}.
\newblock \bibinfo{journal}{\emph{arXiv preprint arXiv:1909.02800}}
  (\bibinfo{year}{2019}).
\newblock
\showeprint[arxiv]{1909.02800}
\urldef\tempurl%
\url{http://arxiv.org/abs/1909.02800}
\showURL{%
\tempurl}


\bibitem[\protect\citeauthoryear{Ram{\'{i}}rez, Krivosheev, B{\'{a}}ez, Casati,
  and Benatallah}{Ram{\'{i}}rez et~al\mbox{.}}{2018}]%
        {CrowdRev2018}
\bibfield{author}{\bibinfo{person}{Jorge Ram{\'{i}}rez},
  \bibinfo{person}{Evgeny Krivosheev}, \bibinfo{person}{Marcos B{\'{a}}ez},
  \bibinfo{person}{Fabio Casati}, {and} \bibinfo{person}{Boualem Benatallah}.}
  \bibinfo{year}{2018}\natexlab{}.
\newblock \showarticletitle{CrowdRev: {A} platform for Crowd-based Screening of
  Literature Reviews}. In \bibinfo{booktitle}{\emph{Collective Intelligence,
  {CI} 2018}}.
\newblock


\bibitem[\protect\citeauthoryear{Rekatsinas, Deshpande, and
  Parameswaran}{Rekatsinas et~al\mbox{.}}{2019}]%
        {DBLP:conf/cikm/RekatsinasDP19}
\bibfield{author}{\bibinfo{person}{Theodoros Rekatsinas}, \bibinfo{person}{Amol
  Deshpande}, {and} \bibinfo{person}{Aditya~G. Parameswaran}.}
  \bibinfo{year}{2019}\natexlab{}.
\newblock \showarticletitle{{CRUX:} Adaptive Querying for Efficient
  Crowdsourced Data Extraction}. In \bibinfo{booktitle}{\emph{{CIKM} 2019}}.
  \bibinfo{pages}{841--850}.
\newblock
\urldef\tempurl%
\url{https://doi.org/10.1145/3357384.3357976}
\showDOI{\tempurl}


\bibitem[\protect\citeauthoryear{Retelny, Bernstein, and Valentine}{Retelny
  et~al\mbox{.}}{2017}]%
        {DBLP:journals/pacmhci/RetelnyBV17}
\bibfield{author}{\bibinfo{person}{Daniela Retelny},
  \bibinfo{person}{Michael~S. Bernstein}, {and} \bibinfo{person}{Melissa~A.
  Valentine}.} \bibinfo{year}{2017}\natexlab{}.
\newblock \showarticletitle{No Workflow Can Ever Be Enough: How Crowdsourcing
  Workflows Constrain Complex Work}.
\newblock \bibinfo{journal}{\emph{{PACMHCI}}} \bibinfo{volume}{1},
  \bibinfo{number}{{CSCW}} (\bibinfo{year}{2017}),
  \bibinfo{pages}{89:1--89:23}.
\newblock
\urldef\tempurl%
\url{https://doi.org/10.1145/3134724}
\showDOI{\tempurl}


\bibitem[\protect\citeauthoryear{Sanh, Debut, Chaumond, and Wolf}{Sanh
  et~al\mbox{.}}{2019}]%
        {DistilBERT2019}
\bibfield{author}{\bibinfo{person}{Victor Sanh}, \bibinfo{person}{Lysandre
  Debut}, \bibinfo{person}{Julien Chaumond}, {and} \bibinfo{person}{Thomas
  Wolf}.} \bibinfo{year}{2019}\natexlab{}.
\newblock \showarticletitle{DistilBERT, a distilled version of {BERT:} smaller,
  faster, cheaper and lighter}.
\newblock \bibinfo{journal}{\emph{arXiv preprint arXiv:1910.01108}}
  (\bibinfo{year}{2019}).
\newblock
\urldef\tempurl%
\url{http://arxiv.org/abs/1910.01108}
\showURL{%
\tempurl}


\bibitem[\protect\citeauthoryear{Shomir and et~al.}{Shomir and et~al.}{2016}]%
        {Wilson2016WWW}
\bibfield{author}{\bibinfo{person}{Wilson Shomir} {and} \bibinfo{person}{et
  al.}} \bibinfo{year}{2016}\natexlab{}.
\newblock \showarticletitle{Crowdsourcing Annotations for Websites' Privacy
  Policies: Can It Really Work?}. In \bibinfo{booktitle}{\emph{WWW 2018}}.
\newblock


\bibitem[\protect\citeauthoryear{Sun, Cheng, Wang, Lyu, Lease, Marshall, and
  Wallace}{Sun et~al\mbox{.}}{2016}]%
        {DBLP:journals/corr/SunCWLLMW16}
\bibfield{author}{\bibinfo{person}{Yalin Sun}, \bibinfo{person}{Pengxiang
  Cheng}, \bibinfo{person}{Shengwei Wang}, \bibinfo{person}{Hao Lyu},
  \bibinfo{person}{Matthew Lease}, \bibinfo{person}{Iain~James Marshall}, {and}
  \bibinfo{person}{Byron~C. Wallace}.} \bibinfo{year}{2016}\natexlab{}.
\newblock \showarticletitle{Crowdsourcing Information Extraction for Biomedical
  Systematic Reviews}.
\newblock \bibinfo{journal}{\emph{arXiv preprint arXiv:1609.01017}}
  (\bibinfo{year}{2016}).
\newblock


\bibitem[\protect\citeauthoryear{Wallace, Noel{-}Storr, Marshall, Cohen,
  Smalheiser, and Thomas}{Wallace et~al\mbox{.}}{2017}]%
        {DBLP:journals/jamia/WallaceNMCST17}
\bibfield{author}{\bibinfo{person}{Byron~C. Wallace}, \bibinfo{person}{Anna
  Noel{-}Storr}, \bibinfo{person}{Iain~James Marshall},
  \bibinfo{person}{Aaron~M. Cohen}, \bibinfo{person}{Neil~R. Smalheiser}, {and}
  \bibinfo{person}{James Thomas}.} \bibinfo{year}{2017}\natexlab{}.
\newblock \showarticletitle{Identifying reports of randomized controlled trials
  (RCTs) via a hybrid machine learning and crowdsourcing approach}.
\newblock \bibinfo{journal}{\emph{{JAMIA}}} \bibinfo{volume}{24},
  \bibinfo{number}{6} (\bibinfo{year}{2017}).
\newblock
\urldef\tempurl%
\url{https://doi.org/10.1093/jamia/ocx053}
\showDOI{\tempurl}


\bibitem[\protect\citeauthoryear{Weiss}{Weiss}{2016}]%
        {Weiss2016CrowdsourcingLR}
\bibfield{author}{\bibinfo{person}{M. Weiss}.} \bibinfo{year}{2016}\natexlab{}.
\newblock \showarticletitle{Crowdsourcing Literature Reviews in New Domains}.
\newblock \bibinfo{journal}{\emph{Technology Innovation Management Review}}
  \bibinfo{volume}{6} (\bibinfo{year}{2016}), \bibinfo{pages}{5--14}.
\newblock


\bibitem[\protect\citeauthoryear{Weng, Li, Hu, and Feng}{Weng
  et~al\mbox{.}}{2017}]%
        {DBLP:conf/cikm/WengLHF17}
\bibfield{author}{\bibinfo{person}{Xueping Weng}, \bibinfo{person}{Guoliang
  Li}, \bibinfo{person}{Huiqi Hu}, {and} \bibinfo{person}{Jianhua Feng}.}
  \bibinfo{year}{2017}\natexlab{}.
\newblock \showarticletitle{Crowdsourced Selection on Multi-Attribute Data}. In
  \bibinfo{booktitle}{\emph{{CIKM} 2017}}. \bibinfo{pages}{307--316}.
\newblock
\urldef\tempurl%
\url{https://doi.org/10.1145/3132847.3132891}
\showDOI{\tempurl}


\bibitem[\protect\citeauthoryear{Whiting, Hugh, and Bernstein}{Whiting
  et~al\mbox{.}}{2019}]%
        {Whiting2019FairWC}
\bibfield{author}{\bibinfo{person}{Mark~E. Whiting}, \bibinfo{person}{Grant
  Hugh}, {and} \bibinfo{person}{Michael~S. Bernstein}.}
  \bibinfo{year}{2019}\natexlab{}.
\newblock \showarticletitle{Fair Work: Crowd Work Minimum Wage with One Line of
  Code}. In \bibinfo{booktitle}{\emph{{HCOMP} 2019}}.
\newblock


\bibitem[\protect\citeauthoryear{Wu and Quinn}{Wu and Quinn}{2017}]%
        {DBLP:conf/hcomp/WuQ17}
\bibfield{author}{\bibinfo{person}{Meng{-}Han Wu} {and}
  \bibinfo{person}{Alexander~J. Quinn}.} \bibinfo{year}{2017}\natexlab{}.
\newblock \showarticletitle{Confusing the Crowd: Task Instruction Quality on
  Amazon Mechanical Turk}. In \bibinfo{booktitle}{\emph{{HCOMP} 2017}}.
\newblock
\urldef\tempurl%
\url{https://aaai.org/ocs/index.php/HCOMP/HCOMP17/paper/view/15943}
\showURL{%
\tempurl}


\bibitem[\protect\citeauthoryear{Wulczyn, Thain, and Dixon}{Wulczyn
  et~al\mbox{.}}{2017}]%
        {WikiDetoxDataset}
\bibfield{author}{\bibinfo{person}{Ellery Wulczyn}, \bibinfo{person}{Nithum
  Thain}, {and} \bibinfo{person}{Lucas Dixon}.}
  \bibinfo{year}{2017}\natexlab{}.
\newblock \showarticletitle{{Wikipedia Talk Labels: Personal Attacks}}.
\newblock  (\bibinfo{date}{2} \bibinfo{year}{2017}).
\newblock
\urldef\tempurl%
\url{https://doi.org/10.6084/m9.figshare.4054689.v6}
\showDOI{\tempurl}


\bibitem[\protect\citeauthoryear{Yang, Redi, Demartini, and Bozzon}{Yang
  et~al\mbox{.}}{2016}]%
        {YangHCOMP2016}
\bibfield{author}{\bibinfo{person}{Jie Yang}, \bibinfo{person}{Judith Redi},
  \bibinfo{person}{Gianluca Demartini}, {and} \bibinfo{person}{Alessandro
  Bozzon}.} \bibinfo{year}{2016}\natexlab{}.
\newblock \showarticletitle{Modeling Task Complexity in Crowdsourcing}. In
  \bibinfo{booktitle}{\emph{{HCOMP} 2016}}.
\newblock


\end{thebibliography}


\end{document}